\documentclass[aps,twocolumn,showpacs]{revtex4}
\usepackage{amssymb}
\usepackage{amsmath}
\usepackage{epsfig}

\begin{document}
\title{A singular integrable equation from short capillary-gravity waves}
\author{M. A. Manna$^{a,b}$ and A. Neveu$^{a,c}$} 
\affiliation{Physique Math\'ematique et Th\'eorique$^a$, CNRS-UMR5825 \\
Universit\'e  Montpellier II, 34095 Montpellier France\\
and\\
Instituto de F\'{\i}sica Te\'orica-UNESP$^b$\\
Rua Pamplona 145, 01405-900 S\~ao Paulo Brazil\\
and\\
Centre de Recherches sur les Tr\`es Basses Temp\'eratures$^c$\\
CNRS-UPR5001, 25 Avenue des Martyrs, BP 166, 38042 Grenoble France }


\begin{abstract} From a columnar approximation of the Euler equations of an
incompressible fluid with surface tension, we derive in the short-wave
approximation a new integrable classical 1+1 dimensional field theory for the
motion of the surface. Together with a Lorentz invariance, 
this system has the novel feature of solutions which become
multiple valued in finite time.
\end{abstract}
\maketitle

\paragraph*{Introduction.} The nonlinear and  dispersive propagation of 
surface waves in an ideal incompressible fluid (depth $h$, density 
$\sigma$), under the action of gravity $g$ and surface tension $T$, is 
a classical subject of investigation in mathematical physics 
\cite{Hunter,Dias,Whitham}. In this letter we derive
and study a new integrable model equation from asymptotic dynamics of 
a short capillary-gravity wave, namely 
\begin{equation} \label{principal} u_{xt} =
\frac{3g(1 - 3\theta)}{2vh}u - \frac{1}{2}u_{xx}u - \frac{1}{4}u_x^2 
+\frac{3h^2}{4v}u_{xx}u_x^2.  \end{equation}
 Here $u(x,t)$ is the fluid velocity on the surface, $x$ and $t$ are space 
and time variables, subindices mean partial derivatives, $\theta = 
(T/\sigma h^2 g$) is the dimensionless Bond number and $v = (3T/\sigma h 
)^{1/2}$.

The dynamics of surface waves in an ideal fluid obeys complicated nonlinear 
and dispersive equations. To simplify them, multiscale asymptotic methods 
can be employed. Most of the resultant asymptotic models represent, for 
large $t$, balance between weak nonlinearities and linear dispersion. For 
instance the long-wave dynamics of a low amplitude initial profile on a 
shallow dispersive fluid are well known nowadays. The models extend from 
the oldest Boussinesq systems or the ubiquitous Korteweg-deVries 
\cite{Whitham} to the more recent Camassa-Holm equation \cite{ch} with 
{\em nonlinear} dispersion. In contrast, hardly anything is known about
asymptotic models for nonlinear and dispersive dynamics of 
{\it short-waves}.  For the most part short waves have been studied 
in connection with  modulation of short-wave trains 
\cite{LHS1,LHS2,Hogan,Mei,ZM}.

In this paper we derive the model \eqref{principal} in the short-wave regime
of surface waves, we prove that it is integrable and show that it leads to
unusual  special solutions that develop singular behavior in finite 
time. 
\paragraph*{The short-wave limit.} 

To define a short wave (wave length $l$, wave number $k=2\pi /l$) one needs 
to compare $l$  to an underlying space scale. We use the unperturbed depth 
$h$  as the natural reference, and thus consider $h= {\cal O}(1)$ and
\begin{equation} \label{limit} kh ={\cal O}(1/\epsilon),\end{equation} 
where $\epsilon$ is the parameter of the asymtotic expansions.

Multiscale asymptotic methods are strongly based on the dispersion relation
$\omega(k)$ and on the associated  phase velocity $v_p$ and group velocity
$v_g$. The short-wave limit \eqref{limit} is meaningful if and only if those
two velocities possess a finite limit. Then $v_p$ and $v_g$ allow to
define asymptotic variables and to handle the nonlinear regime 
\cite{man1,man2}. 

For the usual linearization of the Euler equations (with surface tension),
the linear dispersion relation 
\begin{equation}\label{disp-euler}
\omega(k) = [k(g + \frac{Tk^2}{\sigma})\tanh(kh)]^{1/2}\end{equation} 
yields in the short-wave limit $v_p\sim (Tk/\sigma)^{1/2} \rightarrow 
\infty$. This not only prevents us from defining asymptotic variables
but also infinite dispersion cannot be compensated by weak nonlinearities. 
We found that the solution to this problem is to employ the Green-Nagdhi
conditions of linearization.

\paragraph*{The basic model.}

Green, Laws and Nagdhi \cite{GLN1,GLN2,GLN3} developed alternative 
reductions of the Euler equations leading to models having  dispersion 
relations with good behavior in the short-wave limit, as demonstrated 
in \cite{man1}.  They used three main hypothesis, namely non irrotational
fluid flow, motion in vertical columns and non-Archimedian pressure 
condition. For seek of completeness, we derive hereafter the model 
in a simple manner and include surface tension.

Let the particles of the fluid be identified in a fixed rectangular 
Cartesian system of center $O$ and axes $(x, y, z)$ with  $Oz$ the 
upward vertical direction. We assume translational symmetry in $y$ 
and we will only consider a sheet of fluid in the $xz$ plane. This 
fluid sheet is moving on a rigid bottom at $z = 0$ and its upper 
free surface is $z = S(x,t)$.  The continuity equation and the Newton 
equations (in the flow domain)  read
\begin{align}
& u_x + w_z = 0,\label{C}\\
 & \sigma (u_t + uu_x + wu_z) = -p^*_x,\label{N1}\\
&\sigma ( w_t + uw_x + ww_z) = -p^*_z - g\sigma \label{N2}
\end{align}
where $p^*(x, z, t)$ is the pressure and $(u,w)$ the vectorial velocity. 

The kinematic and dynamic boundary conditions read
\begin{align}
&w = 0\quad {\rm at}\quad z =0\ ,\label{bott}\\
&S_t + uS_x - w  = 0 \quad {\rm at}\quad z = S(x,t),\label{kincond}\\
&p^* = p_0 - \frac{T S_{xx}}{(1 + S^2_x)^{\frac{3}{2}}} \quad {\rm at}
\quad  z = S(x,t).\label{conp}
\end{align}  
The {\it columnar-flow} hypothesis consists in assuming
that $u$ does not dependent on $z$, hence from \eqref{C} and \eqref{bott}
\begin{equation}\label{column}
 u= u(x, t)\ ,\quad  w = -zu_x\ .\end{equation}
The integration
of \eqref{N1} over $z$ from $0$ to $S (x,t)$ then gives
\begin{align}
\sigma S (u_t + uu_x ) = -p_x + T[(1 + S_{x}^2)^{-\frac{1}{2}}]_x,
\label{integrationewton1}\\
p(x, t) = \int_0^{S(x, t)}p^*(x, z, t)dz - p_0 S(x, t).
\end{align}
Now we multiply \eqref{N2} by $z$ and integrate it over $z$ to get
\begin{equation}\label{integrationewton2}
\sigma \frac{S^3}{3}(-u_{xt} - uu_{xx} + u_{x}^2 ) = p + \frac{T S S_{xx}}
{(1 + S_{x}^2)^{\frac{3}{2}}} -\frac{g\sigma S^2}{2}.\end{equation}
Finally, elimination of $p$ between (\ref{integrationewton1})
and (\ref{integrationewton2}) gives, together  with \eqref{kincond} and 
\eqref{column}, the extension of the Green-Nagdhi system to non-zero
surface tension
\begin{align}
S(u_t + uu_x) =&\frac{1}{3}\left[S^3(u_{xt} + uu_{xx} - u_x^2)\right]_x -
g S S_x +\nonumber \\
                   & (T/\sigma) S\ \left[S_{xx}(1+S_x^2)^{-3/2}\right]_x,
\label{M1}\\
S_t + (uS)_x &= 0\ .\label{M2}
\end{align}
This constitutes our basic model.

\paragraph*{Asymptotic model for short capillary-gravity waves.}

In contrast with shallow water theories with dispersion (Boussinesq type
equations), this model incorporates finite dispersion both in
the long-wave and in the short-wave limits.  
Indeed the linear dispersion relation is 
\begin{equation}\label{GN}
\Omega^2 = k^2 [gh +(Th/ \sigma)k^2]/[1 +  (hk)^2/3 ].\end{equation}
Hence the phase velocity  is bounded in the short-wave limit as
we have
\begin{equation} 
\frac\Omega k\sim \left(\frac{3T}{\sigma h}\right)^{1/2} + 
{\cal O}\left(\frac1{h^2k^2}\right).\end{equation} 

This allows to define asymptotic variables \cite{man1}
\begin{equation}\zeta = (1/\epsilon)(x - vt) \ ,\quad
\tau = \epsilon t .\end{equation} 
With the  power series $u = \epsilon^2 (u_0 + \epsilon^2
u_2 + \ldots)$ and $S = h + \epsilon^2 (S_0 + \epsilon^2 S_2 + \ldots)$,
the basic system \eqref{M1}\eqref{M2} leads to
an equation for $u_0(\zeta,\tau)$ which, in the laboratory variable, 
becomes our main equation (\ref{principal}).

\paragraph*{Lax pair and finite-time singularities.} 

After appropriate rescalings of the variables, one can bring equation
(\ref{principal}) into the form
\begin{equation}\label{motion}
u_{xt}=u-uu_{xx}-\frac{1}{2}u^{2}_{x}+\frac{\lambda }{2}u_{xx}u^{2}_{x},
\end{equation}
$\lambda$ being expressed in terms of the physical parameters
of equation (\ref{principal}).
The corresponding Lagrangian  is: 
\begin{equation}\label{lagrangien}
{\cal L}=\frac{1}{2}u_{x}u_{t}+\frac{1}{2}u^{2}+\frac{1}{2}uu^{2}_{x}
-\frac{\lambda }{24}u^{4}_{x}.
\end{equation}

Equation (\ref{motion}) is integrable with Lax pair (in usual notations):
\begin{align}
L=&\frac{\partial }{\partial x}+i \sqrt{E} F \sigma_{3}+ \frac{1}{2}
\frac{u_{xxx}\sqrt{1-\lambda }}{F^{2}}\sigma_{1},\label{LLax}\\
M =&-\frac{1}{2}\left( u-\frac{1}{2}\lambda u^{2}_{x}\right)
\frac{u_{xxx}\sqrt{1-\lambda }}{F^{2}}\sigma _{1}
-i\sqrt{E}\nonumber\\
& \left(u-\frac{1}{2}\lambda u^{2}_{x}\right) F\sigma _{3}
-\frac{i}{4\sqrt{E}}\frac{1-u_{xx}}{F}\sigma _{3}+
\nonumber\\
&  \frac{1}{4\sqrt{E}}
\frac{u_{xx}\sqrt{1-\lambda }}{F}\sigma _{2},\label{MLax}
\end{align}
where $\sigma$ are the usual Pauli matrices, $E$ the ``eigenvalue'' 
and 
\begin{equation}
F^{2}=1-2u_{xx}+\lambda u_{xx}^{2}.
\end{equation}
One of its most remarkable properties is that with 
$F$ one builds the first non-trivial conserved quantity for all $\lambda$:
\begin{equation}
F_t = \left[\left(u-\frac{\lambda}{2}u_x^2\right)
F\right]_x,
\end{equation}
and through the change of function from $u(x,t)$ to
\begin{equation}
\label{changefunc}
g(y,t)=\frac{1}{\sqrt{1-\lambda}}{\rm Argtanh}\frac{u_{xx}
\sqrt{1-\lambda}}{1-u_{xx}},
\end{equation}
with 
\begin{equation}
y=\int^x Fdx,
\end{equation}
one finds that $g$ satisfies the sinh-Gordon equation
\begin{equation}
\label{sinh}
g_{yt} =\frac{1}{\sqrt{1-\lambda}}{\rm sinh}\sqrt{1-\lambda}g.
\end{equation}

This is valid for $\lambda<1$ and for $u_{xx}$ small enough so that $F$ is
real. If $u_{xx}$ is large, a similar change leads to the cosh-Gordon 
equation, and if  $\lambda>1$ one obtains the sine-Gordon equation. Finally,
for $\lambda=1$, one obtains for $g(y,t)=u_{xx}/(1-u_{xx})$ the equation of 
a free field in light-cone coordinates.

Whatever the value of $\lambda$, it follows from the change of variables 
$(x,t)\leftrightarrow (y,t)$ that a regular $g(y,t)$ can give back a 
{\em singular, multivalued} $u(x,t)$ if the change from $y$ back to $x$ 
is not one-to-one. This happens when $|g|$ is large enough, forcing 
$u_{xx}$ to infinity and a change of sign of $F$ in the equation for $y$.
We give an example of this in fig. 1, where two solutions for $u$ are 
plotted from breather solutions of the sine-Gordon equation, one 
(dashed curve) with an amplitude just below the singularity threshold,  
so that  $u$ is still regular and single-valued, the other (solid curve)  
 with an amplitude above  the singularity threshold, which displays a 
swallowtail behavior.

\begin{figure}[ht]
\centerline{\epsfig{file=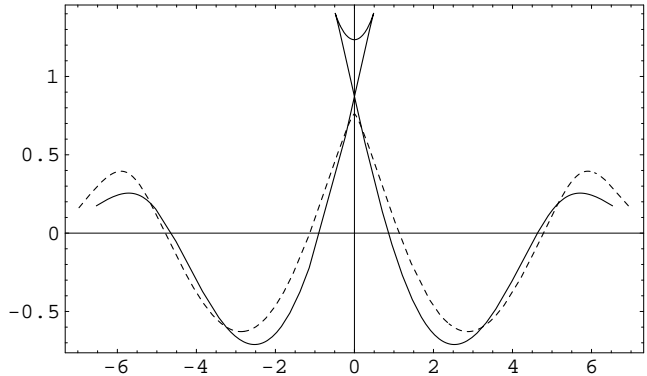,height=5cm,width=8.5cm}}
\caption{$u(x,t=0)$ corresponding to sine-Gordon breathers
$g(y,t)\sqrt{\lambda-1}=-4{\rm Arctan}\frac{d\cos[c(y-t)]}{\cosh[s(y+t)]}, 
c^2+s^2=1, d=s/c,$ 
at $\lambda=2.02,$ $d=0.2$ (dotted curve),
 $d=0.35$ (solid curve).}
\label{fig:ubreather}\end{figure}

Furthermore, this singularity and multivaluedness can be reached
in finite time from a regular solution.
In particular, in the sine-Gordon case, one can start from a solution 
$g(y,t)$ consisting of two breathers far enough from each other and
each weak enough so that the change $y \rightarrow x$ is single valued,
but strong enough so that when they overlap, $|g|$ becomes large enough
for the singularity to appear. In the free field case $\lambda=1$,
and even in the sinh-Gordon case,
one can replace the breathers with wave packets, and make them 
collide and give rise to this singularity before they disperse.
In all these cases, the singularity in $u$ is a swallowtail
just like in figure 1.

The (singular, multivalued) $u(x,t)$ corresponding to 
the sine-Gordon soliton for $g(y,t)$ is also interesting.
It is displayed on figure 2 for the particular value $\lambda = 10/9$.

\begin{figure}[ht]
\centerline{\epsfig{file=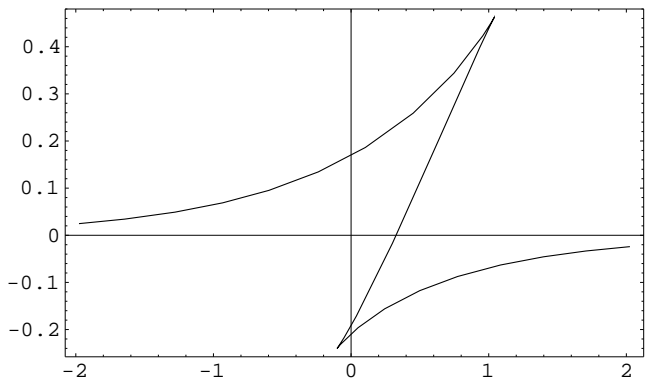,height=5cm,width=8.5cm}}
\caption{$u(x,t=0)$ corresponding to the sine-Gordon soliton
$g(y,t)\sqrt{\lambda-1}=4{\rm Arctan} \exp (y+t)$ at  $\lambda=10/9$.}
\label{fig:usoliton}\end{figure}

For $\lambda=0 $,  equation (\ref{motion}) was already discussed in
\cite{kruskal,cewen,hunter2,alber1,alber2,alber3}, and it 
contains peakons.
These peakons, which are solitons, and their scattering 
are qualitatively very easy to study {\em via} the change
of variables to the sinh-Gordon equation, where  they
correspond to singular solutions obtained by simple
analytic continuations of the sine-Gordon multisoliton solutions.

For any value of $\lambda$, the equation (\ref{motion}) has the symmetry
$x\rightarrow ax, \quad  t\rightarrow t/a, \quad   u\rightarrow a^2u,$
for arbitrary real $a$, which is just the Lorentz group in 1+1 dimensions.
Although the Lagrangian (\ref{lagrangien}) does not have the
appropriate covariance property to give a Lorentz invariant action,
the system, being integrable, has an infinite set of Lagrangians
(and derived Hamiltonians and Poisson brackets) leading to the
same equation of motion (\ref{motion}), and one of them, built
with the invariant field $u_{xx}$, leads to an invariant action.

\paragraph*{Benjamin-Feir instability.} 

The Benjamin-Feir instability results from resonant interaction of an 
initial monochromatic wave with side-band modes produced by nonlinearity.
This instability, which modulates the carrier envelope, is well described
by the nonlinear Schr\"odinger asymptotic limit.

Following the standard approach \cite{Whitham} we can show that  
a Stokes wave train of equation (\ref{principal}) is unstable if 
\begin{equation}
\theta < \frac{3}{10},
\end{equation} 
namely any slight deformation of the
plane wave experiences an exponential growth.
In the case of water at room temperature 
($T=0.074\ Nm^{-1}$, $\sigma=10^3\ kg\ m^{-3} $), 
we obtain that a short wave train is unstable for a depth $h>0.49\ cm$.

Last but not least, the value $\theta=3/10$ corresponds to $\lambda=1$ in
\eqref{motion}. Precisely
\begin{equation}
\theta<0.3\ \Rightarrow\ \lambda>1.\end{equation}

\paragraph*{Comments.} 

The occurence of singular (even multivalued) solutions in an equation 
derived in a hydrodynamics framework is interesting, especially since 
$u$ is the deviation of the free surface from equilibrium. The actual 
observability of the present singular behaviors would require a more 
detailed analysis of the validity of the short-wave approximation near 
the singular points which goes beyond the scope of this paper.  
In particular the inclusion of viscosity, which acts strongly over small
scales, will affect the short-waves dynamics and alter these singular 
behaviors.

The Green-Naghdi equations can be improved systematically toward higher 
depths \cite{Kim}. In the linear limit, these improvements give 
the higher $(N,N)$ Pad\'e approximants of the Euler dispersion relation 
\eqref{disp-euler}, the present case \eqref{GN} corresponding to $N=1$. 
In particular they always lead to a finite phase velocity in the 
short-wave limit, which grows quickly as the order of the approximation 
grows to try to mimic the behavior of the exact formula.

What is remarkable is that the short-wave asymptotics of these improvements
lead to exactly the same integrable equation (\ref{principal}) except for
different numerical coefficients in front of the physical quantities $g$,
$h$, $\theta$ and $v$.  Hence we hope that at least some the main features
of the singular behavior of the solutions correspond to the actual physics
of these water waves in arbitrary depth.  

Another point is the existence of a peakon solution in the case 
$\lambda = 0$. In the rescaling leading from equation (\ref{principal}) 
to equation (\ref{motion}), this value of $\lambda$ is obtained only for 
$\theta=1/3$, where the rescaling is singular for the $x,t$ and $u$ 
variables themselves, so that the whole asymptotics must be reconsidered
from the start. This value $\theta=1/3$ in the Euler  equation leads to a
dispersionless system for small $k$ only and it is a peculiar feature of 
the Green-Naghdi equations to be dispersionless for all $k$ (hence for 
large $k$) for that value of $\theta$. This large $k$ feature is not 
inherited by the improvements of \cite{Kim}.

Equation (\ref{principal}) has a Lorentz invariance, and its quantum 
version promises to exhibit new features not shared by the existing 
relativistic systems in 1+1 dimension.  This new relativistic integrable
system, with just one massive bosonic field and a local classical 
equation of motion, is quite intriguing. In particular, the change 
of variable which transforms it into sine- or sinh-Gordon mixes the 
space-time variable $x$ and the field $u_{xx}$ and furthermore requires 
the equation of motion to be satisfied. Hence, it cannot be quantum 
mechanically equivalent to sine- or sinh-Gordon.  For example, it is not 
parity invariant (parity in the laboratory frame, $x+t\rightarrow -x-t$, 
$x-t\rightarrow x-t$).  From this follows an unusual $S$-matrix, in which
 there are two phase-shifts, one for the left moving particle and one for
 the right moving one. The quantum field theory and mathematical 
structures following from this are worth a detailed study for themselves 
in separate publications. This could also have some analogy in
general relativity, where space-time is, through the metric, a dynamical
variable and where one can go from a choice of space-time parametrization
to another one by a change which can involve the metric itself.

\paragraph*{Acknowledgements.} This work was supported in part by FAPESP
(Funda\c c\~ao de Amparo \`a Pesquisa do Estado de S\~ao Paulo) and 
Research Training Network grant of the European Commission contract number
HPRN-CT-2002-00325. M. A. M. wish to thank R. A. Kraenkel and J. L\'eon for
stimulating discussions and IFT for hospitality. A. N. is grateful to A. V.
Mikhailov and V. I. Zakharov for discussions and to the Newton Institute 
for its hospitality.

\end{document}